\documentstyle[12pt,aaspp4]{article}
\begin{document}

\title{Extracting Energy from Black Hole through Transition Region}

\author{Li-Xin Li}
\affil{Princeton University Observatory, Princeton, NJ 08544--1001, USA}
\affil{E-mail: lxl@astro.princeton.edu}

\begin{abstract}
A new scenario for extracting energy from a Kerr black hole is proposed.
With magnetic field lines connecting plasma particles inside the ergosphere
with remote loads, the frame dragging twists the field lines so that energy and
angular momentum are extracted from the plasma particles. If the magnetic
field is strong enough, the energy extracted
from the particles can be so large that the particles have negative energy
as they fall into the black hole. So effectively the energy is extracted from
the black hole. 

The particles inside the ergosphere can be continuously replenished with 
accretion from a disk surrounding the black hole, so a transition region 
with sufficient amount of plasma is formed between the black hole's horizon
and the inner edge of the disk. Thus the energy can be continuously extracted 
from the black hole through the transition region. This may be the most
efficient way for extracting energy from a Kerr black hole: in principle
almost all of the rotational energy (up to $\approx 29\%$ of the total 
energy of the black hole) can be extracted.
\end{abstract}


\section{Introduction}
How to extract energy from a rotating black hole is an interesting and
important issue in astrophysics. For a Kerr black hole, up to $\approx
29\%$ of its total energy (i.e. all of its rotational energy) is available for
extraction in principle (Christodoulou 1970). Penrose (1969) was the first to
propose a Gedankenexperiment for extracting energy from a Kerr black hole
by splitting a particle into two inside the ergosphere: one particle with
negative energy (relative to observers at infinity) is captured by the black
hole, the other with energy greater than that of the original particle escapes
to infinity. However, the Penrose process has been argued to be impractical
in astrophysics since it requires the two new-born particles separate from each
other with speed greater than half of the light speed (Bardeen, Press, \&
Teukolsky 1972; Wald 1974). Since then many people have considered various
alternative mechanisms for extracting energy from a rotating black hole.
Among them the most promising one is the Blandford-Znajek
mechanism (Blandford \& Znajek 1977), which in fact has been proposed to be
a possible process for powering jets in quasars and active galactic nuclei 
(Rees, Begelman, Blandford, \& Phinney 1982; Begelman, Blandford, \& Rees 1984)
and gamma-ray bursts (Paczy\'nski 1993; M\'esz\'aros \& Rees 1997; Lee, Wijers, 
\& Brown 2000). In the Blandford-Znajek mechanism, a Kerr black hole is assumed 
to connect with remote loads with magnetic field lines. The magnetic field lines 
thread the black hole's horizon, the rotation of the black hole twists the
magnetic field lines and transports energy and angular momentum from
the black hole to the loads (Blandford \& Znajek 1977; Macdonald \& Thorne
1982; Phinney 1983). Recently, a variant picture has been
considered: instead of connecting with remote loads, the magnetic field
lines are assumed to connect a Kerr black hole with a disk surrounding it
(Blandford 1999; Gruzinov 1999; Li 2000). If the black hole rotates faster
than the disk, energy and angular momentum are extracted from the black hole
and transferred to the disk and ultimately radiated away by the disk
(Li 2000).

In this paper we consider a new scenario for extracting energy from a Kerr
black hole. Assume some magnetic field lines connect plasma particles inside the
ergosphere of a Kerr black hole with remote loads. Because of the frame dragging
of the Kerr black hole, plasma fluid inside the ergosphere always has a positive
angular velocity relative to observers at infinity (``positive" means that the 
angular velocity of the plasma particles and the angular velocity of the black 
hole have the same sign). Because of the rotation of the plasma fluid,  energy 
and angular momentum are extracted from the plasma particles
by the magnetic field lines and transported to the remote loads.
If the magnetic field is strong enough, so much energy can be extracted that
as the particles fall into the horizon they have negative energy relative to
observers at infinity. So, effectively, the energy is extracted from the
black hole. The particles inside the ergosphere can be continuously 
replenished with accretion from  a disk surrounding the black hole, so a 
transition region with sufficient amount of plasma particles is formed
between the black hole's horizon and the inner edge of the disk. Thus energy
can be continuously extracted from the black hole through the transition
region. We will show that this may be the most efficient way for 
extracting energy from a Kerr black hole: in principle almost all of its 
rotational energy (up to $\approx 29\%$ of its total energy) can be extracted.

Similar processes inside the ergosphere have been considered by other people,
who have assumed that magnetic field lines connect an accretion disk either with 
particles in the transition region (Gammie 1999; Agol \& Krolik 2000; however 
see the critical comments of Paczy\'nski 2000, and Armitage, Reynolds, \& Chiang 2000) 
or with remote loads (Meier
1999). Then the accreting particles can fall into the black hole with negative
energy and negative angular momentum if the magnetic coupling is strong enough, 
so energy and angular momentum flow from the black
hole into the disk. While in our scenario, the magnetic field lines connect
particles in the transition region with remote loads without going through
the disk. The energy extracted from the black hole is transported to infinity
through the transition region instead of going through the disk.

\section{Extraction of Energy and Angular Momentum}
Consider some magnetic field lines connecting plasma fluid inside the ergosphere
of a Kerr black hole with remote loads. The foot-points of the magnetic field
lines in the plasma fluid are supposed to be close to the event horizon of the
black hole, where the angular velocity of the fluid is $\approx \Omega_{\rm H}$ 
because of the frame dragging, where $\Omega_{\rm H}$ is the angular velocity of 
the black hole (Misner, Thorne, \& Wheeler 1973). The electromotive force induced 
in the plasma fluid by the frame dragging is ${\cal E} \approx \Omega_{\rm H}
\Delta\Psi/2\pi$, where $\Delta\Psi$ is the magnetic flux threading the plasma.
We use the geometric units with $G = c = 1$ throughout the paper.
The power on the loads (i.e. the energy extracted per unit time from the fluid
in the transition region, which is measured by observers at infinity) is
\begin{eqnarray}
   P \approx \left({\Delta\Psi\over 2\pi}\right)^2\, {\Omega_{\rm H}^2 
             Z_{\rm L}\over
             \left(Z_{\rm L} + Z_0\right)^2}
     = \left({\Delta\Psi\over 2\pi}\right)^2\, {\Omega_{\rm F}
       \left(\Omega_{\rm H} -
       \Omega_{\rm F}\right)\over Z_0}\,,
   \label{pow}
\end{eqnarray}
where $Z_{\rm L}$ is the resistance of the loads, $Z_0$ is the resistance of the
plasma fluid, and
\begin{eqnarray}
   \Omega_{\rm F} = {\Omega_{\rm H} Z_{\rm L}\over Z_{\rm L} + Z_0}
\end{eqnarray}
is the angular velocity of
the magnetic field lines. The torque on the loads (i.e. the angular momentum 
extracted per unit time from the fluid in the transition region, which is 
measured by observers at infinity) is $T = P/\Omega_{\rm F}$. If the plasma
inside the ergosphere is perfectly conducting so $Z_0$ is small, $P$ can
easily reach a big value for a wide range of $Z_{\rm L}$: from $Z_{\rm L}\approx
Z_0^2/Z_{\rm H}$ to $Z_{\rm L}\approx Z_{\rm H}$, where $Z_{\rm H}$ is the 
resistance of the black hole which is $\gg Z_0$. In other words, we 
do not need ``impedance-matching'' --- which is needed by the Blandford-Znajek
mechanism (Macdonald \& Thorne 1982). 

The plasma particles inside the ergosphere are continuously replenished with
accretion from a disk surrounding the black hole with accretion rate $\dot{M}=
dM/dt$ (measured by observers at infinity). In a steady state, the black hole
must capture particles at the same rate. If the particles captured by the
black hole have negative energy (measured by observers at infinity; we use
the convention $\dot{M}>0$ for accretion), energy is extracted from the black 
hole. In a steady state the power and the torque of the black
hole are respectively
\begin{eqnarray}
   P_{\rm H} = P - \dot{M} e_i\,, \hspace{1 cm}
   T_{\rm H} = T - \dot{M} j_i\,,
   \label{bal}
\end{eqnarray}
where $P$ is given by equation (\ref{pow}), $T = P/\Omega_{\rm F}\approx 
P/\Omega_{\rm H}$ assuming $Z_{\rm L}\gg Z_0$, and $e_i$ is the specific energy 
($0<e_i<1$),  $j_i$ is the 
specific angular momentum, of the particles before they enter the transition 
region. When $P>\dot{M} e_i$, we have $P_{\rm H}>0$, the particles fall into 
the black hole with negative energy, so energy is extracted from the black 
hole\footnote{A particle with negative energy inside the ergosphere must have 
negative angular momentum, so $P_{\rm H}>0$ implies $T_{\rm H}>0$. But the 
reverse is incorrect: a particle with negative angular momentum inside the 
ergosphere does not have to have negative energy, so $T_{\rm H}>0$ does not 
imply $P_{\rm H}>0$. The angular velocity of a particle inside the ergosphere 
is always positive, no matter its energy (or angular momentum) is positive or 
negative.}. However, the angular velocity of the particles 
inside the ergosphere is always positive and so energy is continuously
extracted from the particles and transported to the remote loads until
the particles fall into the black hole. (The fluid-lines of the plasma 
particles in the transition region are illustrated in Fig. \ref{fig1}.) 

For a fast rotating black hole with $a/M_{\rm H}\approx 1$ where 
$M_{\rm H}$ is the mass of the black hole and $a M_{\rm H}$ is the angular 
momentum of the black hole, the magnitude of $P$ can be estimated as $\approx
B^2 r_{\rm H}^2$ if $Z_{\rm L}\approx Z_{\rm H}$ as in the case of the 
Blandford-Znajek mechanism, where $B$ is the strength of the magnetic field 
in the transition
region, $r_{\rm H}$ is the radius of the black hole's horizon. By absorbing all 
ambiguous factors (with order of magnitude $\sim 1$) into $B$ and $\dot{M}$, 
the condition for $P_{\rm H}>0$ is given by $B^2 r_{\rm H}^2 > \dot{M}$, or
\begin{eqnarray}
    {\dot{M}\over \dot{M}_{\rm Edd}} < 0.1 \left({M\over 10^9M_{\odot}}\right)
         \left({B\over 10^4\,{\rm Gauss}}\right)^2\,,
    \label{cond}
\end{eqnarray}
where $\dot{M}_{\rm Edd}\approx 10^{27}\left(M_{\rm H}/10^9 M_{\sun}\right)$ g/s
is the Eddington limit of accretion rate.

\section{Efficiency in Extracting Energy from Black Hole}
Suppose energy and angular momentum are extracted from a Kerr black hole
at rates $P_{\rm H}$ and $T_{\rm H}$ respectively. As the black hole's spin 
$s = a/M_{\rm H}$ goes down from $s_0$ to $s$, the total energy extracted from 
the black hole is $E = \eta M_{\rm H}$, where $M_{\rm H}$ is the black hole's
initial mass, $\eta$ is the efficiency for the process which can be 
calculated by (Li 2000)
\begin{eqnarray}
   \eta = 1 - \exp\int_{s_0}^s {ds\over \left(\mu M_{\rm H} \Omega_{\rm H}
          \right)^{-1} - 2s}\,,
   \label{eff}
\end{eqnarray}
where $\mu = P_{\rm H}/\left(T_{\rm H}\Omega_{\rm H}\right)$ measures
the step-by-step efficiency in extracting energy from the black hole. The value
of $\mu$ depends on the process. For the process described in the previous
section, $\mu\approx \Omega_{\rm F}/\Omega_{\rm H}\approx 1$ with the 
assumptions of $Z_0\ll Z_{\rm L}$ and $P\gg \dot{M} e_i$. For $s_0 = 1$ and 
$s = 0$ we have $\eta \approx 29\%$. Compared with the
Blandford-Znajek mechanism and the magnetic coupling between a black hole and
its disk (Li 2000), the precess presented in this paper is most efficient in 
extracting energy from a Kerr black hole. In principle, the amount of energy 
that can be extracted from a Kerr black hole through the transition region can 
be as close as we want to the total rotational energy, which is up to $\approx 
0.29 M_{\rm H}$ for $a/M_{\rm H} = 1$. The Blandford-Znajek mechanism can extract
$\approx 0.09 M_{\rm H}$, the magnetic coupling between a black hole and its 
disk can extract $\approx 0.16 M_{\rm H}$ for $s_0 = 1$. (See Fig. \ref{fig2}.)

By the first law of black hole mechanics we have
\begin{eqnarray}
   P_{\rm H} = T_{\rm H}\Omega_{\rm H} - {\kappa\over 8\pi}\,{dA\over dt}\,,
\end{eqnarray}
where $\kappa$ is the surface gravity of the black hole, $A$ is the surface 
area of the black hole's horizon. The term $(8\pi)^{-1}\kappa\,
dA/dt$ represents the rate of energy dissipation in the black hole, i.e. the
increase in the black hole's entropy. The second law of black hole
physics says $dA/dt\ge 0$ always. [For the four laws of black hole mechanics
see Bardeen, Carter, \& Hawking (1973).] So, to gain a high efficiency in
extracting energy from a Kerr black hole, $P_{\rm H}/T_{\rm H}$ should be as
close as possible to $\Omega_{\rm H}$, but keep less than $\Omega_{\rm H}$. For 
the process discussed in this paper, with $P\gg \dot{M} e_i$, we have $P_{\rm H}/
T_{\rm H}\approx \Omega_{\rm F}$ which is close to but less than 
$\Omega_{\rm H}$ near the horizon of the black hole. So through the transition
region --- whose inner boundary is at the black hole's horizon --- we can get 
the highest efficiency in extracting energy from a Kerr black hole. 

The black hole spins down as it loses angular momentum. However, it can spin up
again through accretion from the disk (Bardeen 1970). By alternating the spin-up
process associated with accretion and the spin-down process associated with
magnetic braking, an indefinite cyclic process can be constructed with the black
hole's spin $s = a/M_{\rm H}$ oscillating between $s_1$ and $s_2$ (Li \& 
Paczy\'nski 2000). In each cycle the overall efficiency in converting mass 
into energy is
\begin{eqnarray}
   \eta_t = {E_1+E_2\over \Delta M_{\rm D}}\,,
\end{eqnarray}
where $E_1$ is the energy radiated in the spin-up phase, $E_2$ is the energy
radiated in the spin-down phase, and $\Delta M_{\rm D}$ is the mass accreted 
from the disk during the spin-up phase. (In the spin-down phase the mass 
accreted from the disk
is assumed to be negligible.) $E_1$ can be calculated with equations in
Bardeen (1970), $E_2$ can be calculated with $E_2 = \eta M_{\rm H}$ where
$\eta$ is given by equation (\ref{eff}). We find that, $\eta_t$ peaks at $s_1
\approx 0.7288$ and $s_2 \approx 0.99997$ for $0\le s_1< s_2<1$, and
$\eta_{t,\max}\approx 63.8\%$. In each cycle the black hole's mass increases
$\approx 76\%$. If we restrict $s$ in the interval $0\le s\le 0.998$, as
suggested by Thorne (1974)\footnote{However, Abramowicz and Lasota (1980) have
argued that Thorne's limit can be exceeded for a thick accretion disk. And, 
we should
keep in mind that the upper limit on $s$ given by the third law of black hole
physics is 1 (Bardeen, Carter, \& Hawking 1973).}, then $\eta_t$ peaks at
$s_1\approx 0.8076$ and $s_2\approx 0.998$, but $\eta_{t,\max}$ doesn't change
much: $\eta_{t,\max}\approx 63.2\%$. In this case the black hole's mass
increases $\approx 52\%$ per cycle. This efficiency is higher than that in the
case with the black hole spun-down by the magnetic coupling with the disk,
which is $\approx 43.6\%$ (Li \& Paczy\'nski 2000).

In realistic cases, magnetic field lines distribute over the transition
region instead of touching the transition region exactly at the inner
boundary. Then the averaged angular velocity of the magnetic field lines 
$\overline{\Omega}_{\rm F} = P/T \approx P_{\rm H}/T_{\rm H}$ is
smaller than $\Omega_{\rm H}$, i.e. $\mu = P_{\rm H}/\left(T_{\rm H}
\Omega_{\rm H}\right) <1$, thus the total efficiency in extracting energy 
from the black hole is smaller than that estimated above. From equation 
(\ref{eff}), for fixed values of $s$ and $s_0$, $\eta$ increases as $\mu$
increases from $0$ to $1$. The Blandford-Znajek mechanism has $\mu \approx
1/2$ and $\eta \approx 0.09$. Thus, if the averaged value of $\mu$ lies
between $1/2$ and $1$, the mechanism of extracting energy from a black hole
through the transition region has a higher efficiency than the 
Blandford-Znajek mechanism does. Otherwise, the mechanism of extracting energy 
from a black hole through the transition region has a lower efficiency. 
$\eta \approx 29\%$ provides an upper limit for the efficiency in extracting 
energy from a Kerr black hole, as originally shown by Christodoulou (1970).

\section{Conclusions and Discussions}
As plasma particles in a magnetized accretion disk accrete onto a black hole, 
the magnetic field lines frozen in the plasma drift toward the black hole 
together. Since a black hole cannot hold magnetic fields by itself, excess 
magnetic field lines coming with the accreting plasma will be expelled from 
the black hole when the magnetic flux threading the black hole's horizon gets
saturated. So, we can expect that as accretion goes on, in the transition 
region between the horizon of the black hole and the inner boundary of the disk 
a strong magnetic field will be built up near the black hole's horizon. 
This magnetic field could be stronger than both the magnetic field on the black 
hole's horizon and the magnetic field in the accretion disk. Along with the 
accretion of the plasma particles, the toroidal currents in the disk accrete 
onto the black hole. As these currents get into the black hole, they are quickly
dissipated since a black hole is a poor conductor. Accompanying the dissipation
of currents, some magnetic field lines diffuse from the black hole to the 
transition region, which is equivalent to accumulation of toroidal currents
in the transition region. The toroidal currents accumulated in the transition
region sustain the magnetic field in the transition region and the magnetic
field on the black hole's horizon. The 
magnetohydrodynamics inside the transition region is very complicated.
However, in a steady state with low accretion rate, the plasma in the 
transition region is expected to have a density $\rho\sim B^2/8\pi$, which 
provides sufficient amount of plasma particles for carrying poloidal currents 
flowing across the magnetic field lines (Krolik 1999).

For a Kerr black hole, the inner part of the transition region is inside the 
ergosphere. In fact, for a Kerr black hole with a thin Keplerian
disk, the whole transition region is inside the ergosphere when $a/M_{\rm H}>
0.9428$. Because of the frame dragging the plasma fluid inside the ergosphere 
always has a positive angular velocity. If the magnetic field lines 
threading the transition region inside the ergosphere connect with remote 
astrophysical loads, energy and angular momentum can be extracted from the 
transition region by magnetic braking. When the accretion rate is low and 
the magnetic field is strong enough, so much energy can be extracted that the 
plasma particles fall into the black hole's horizon with negative energy.
Then the energy is essentially extracted from the black hole, though the 
magnetic field lines relevant to this process don't touch the black hole at all.

This provides a new scenario for extracting energy from a Kerr black hole
with some interesting features: (1) Since the resistance in the transition
region can be much smaller than that of the black hole, and the angular velocity
of the plasma fluid near the black hole horizon  is close to $\Omega_{\rm H}$, 
the efficiency in extracting energy from the black hole can be very high --- 
in an optimal case it can be close to the limit $\approx 29\%$ for a Kerr black 
hole with initial 
$a/M_{\rm H} = 1$; (2) Since inside the ergosphere particles with negative 
energy must be captured by the black hole, the energy extracted in this way 
is expected to be cleaner than that extracted from the disk; (3) Since 
the transition region has a small resistance, a high power can be gained 
more easily than that in the Blandford-Znajek mechanism.

\acknowledgments{I am very grateful to Bohdan Paczy\'nski and
Jeremy Goodman for helpful and stimulating
discussions. This work was supported by the NASA grant NAG5-7016.}


\newpage
\figcaption[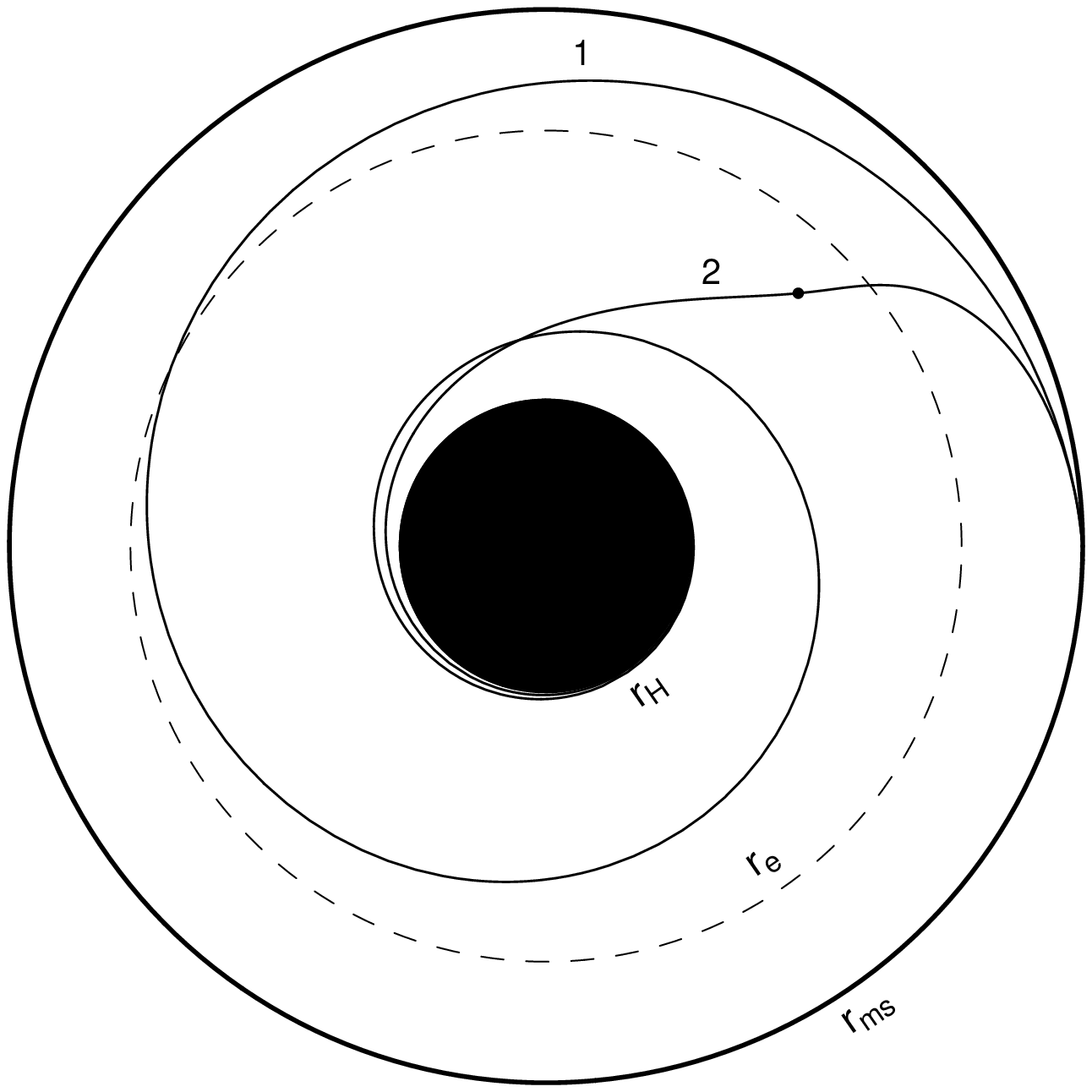]{A sketch of the fluid-lines of the plasma particles
in the transition region around a Kerr black hole. Two different kind of 
fluid-lines in the equatorial plane are shown. When there is no magnetic
field, the accreting particles free-fall into the black hole through the transition
region between the marginally stable orbit (marked as $r_{\rm ms}$, which is
the inner boundary of a thin Keplerian disk) and the black hole horizon (marked as 
$r_{\rm H}$). The orbit of such a particle is shown with the curve marked with ``1''.
The particle 
on this orbit has positive energy and positive angular momentum which are conserved 
along the orbit. For the process discussed in the text --- energy and angular 
momentum are extracted from the particles in the transition region by magnetic 
field lines, the accreting particles lose their energy and angular momentum as 
they spiral towards the black hole. The orbit of such a particle is shown with the 
curve marked with ``2''. The particle leaves the inner boundary of the disk with 
positive energy and positive angular momentum. However, as the particle gets close 
to the horizon so much energy and angular momentum are lost that the particle has 
negative energy and negative angular momentum. The position where the energy of 
the particle is zero is shown as a thick dot on the orbit, which is inside the 
ergosphere --- the region between the horizon and the dashed circle marked with 
$r_{\rm e}$. (The marginally stable orbit is inside the ergosphere when $a/M_{\rm H}
> 0.9428$, where $M_{\rm H}$ and $a M_{\rm H}$ are the mass and the angular momentum 
of the Kerr black hole.) The losing of angular momentum tries to decrease the 
angular velocity of the particle. However, in the neighbor
of a Kerr black hole the frame dragging is so important that the angular velocity of
the particle is always positive. In fact, near the black hole horizon the angular 
velocity of the particle keeps increasing though the particle is losing angular
momentum and energy. As the particle approaches the horizon, its angular velocity
approaches $\Omega_{\rm H}$ --- the angular velocity of the black hole, which
is a general feature independent of the energy and angular momentum of the
particle. (In the diagram the radial coordinate is scaled logrithmly.)
\label{fig1}}

\figcaption[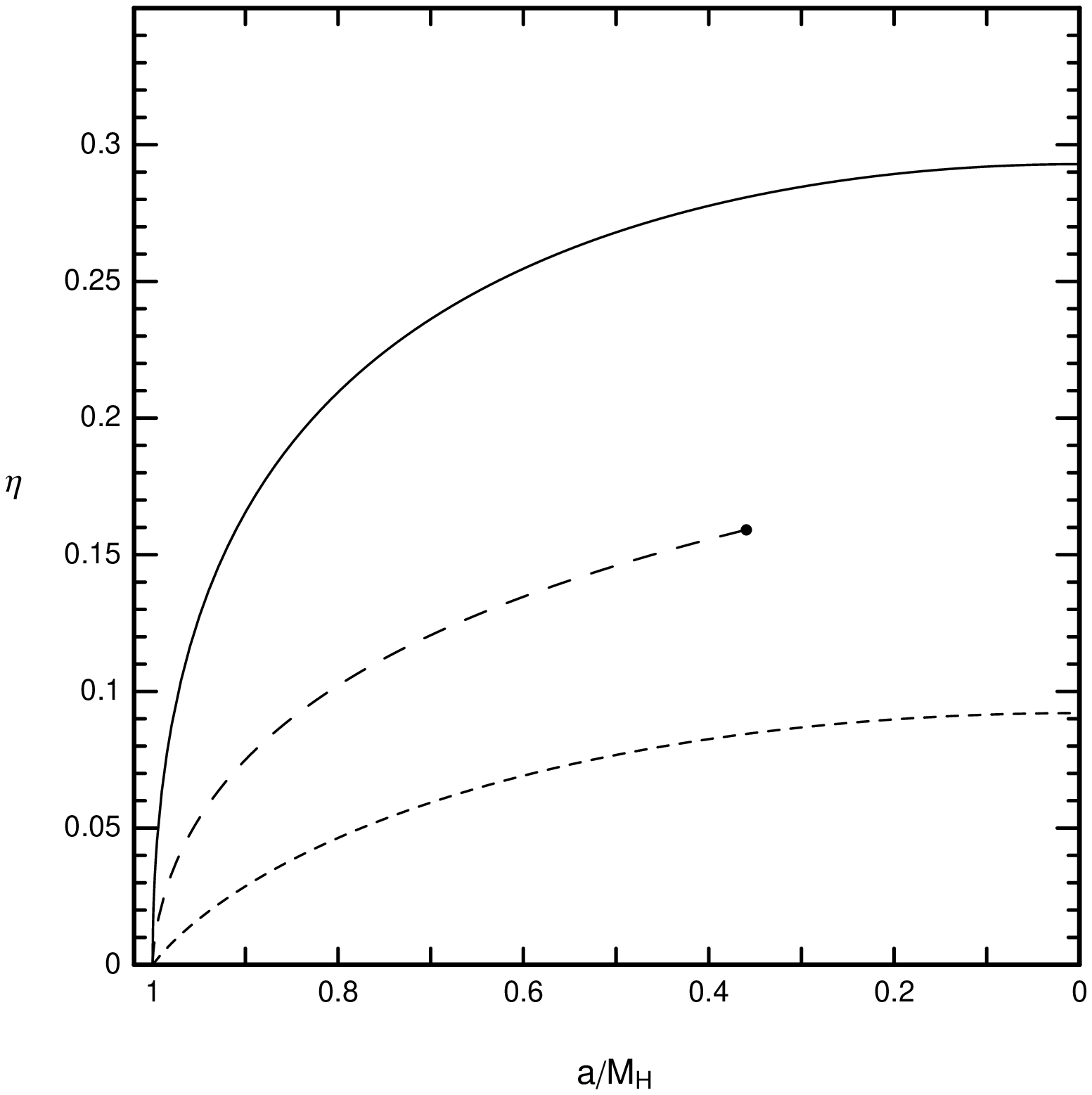]{The efficiency in extracting energy from a Kerr black hole
as the black hole is spun down. The efficiency is defined by $\eta =
E/M_{\rm H}$, where $M_{\rm H}$ is the mass of the black hole at its initial
state with $a/M_{\rm H} = 1$, $E$ is the integrated energy extracted from the 
black hole. Note that in the figure $a/M_{\rm H}$ decreases
from left to right. The solid curve represents the process discussed in the
paper: extracting energy from a Kerr black hole through the transition region. 
The magnetic field lines touch the transition region close to the horizon of the 
black hole, so $\Omega_{\rm F}/\Omega_{\rm H}\approx 1$, where $\Omega_{\rm H}$
is the angular velocity of the black hole, $\Omega_{\rm F}$ is the angular velocity
of the magnetic field lines. This optimal process has the highest efficiency in 
extracting
energy from the black hole: $\eta\approx 29\%$ as $a/M_{\rm H}$ gets $0$, where
$a$ is the specific angular momentum of the black hole. In realistic cases, 
magnetic field lines distribute over the transition region instead of touching 
the transition region exactly at its inner boundary, then the efficiency is smaller 
than that given by the solid curve. Thus, the solid curve provides an upper limit
on the efficiency in extracting energy from a Kerr black hole. 
The long-dashed curve represents the process 
of extracting energy from a Kerr black hole through the magnetic coupling with
a thin Keplerian disk (Li 2000), which ends at $a/M_{\rm H} = 0.3594$
since then the transfer of energy and angular momentum from the black hole to 
the disk stops. In this process $\Omega_{\rm F}/\Omega_{\rm H}$ varies with 
$a/M_{\rm H}$. As $a/M_{\rm H}$ gets  $0.3594$, the efficiency $\eta\approx 16\%$. 
The short-dashed curve represents the Blandford-Znajek mechanism, which has 
$\Omega_{\rm F}/\Omega_{\rm H}\approx 1/2$. As $a/M_{\rm H}$ gets $0$ the 
efficiency of the Blandford-Znajek mechanism is $\eta\approx 9\%$.
\label{fig2}}


\begin{references}

\reference{} Abramowicz, M. A., \& Lasota, J. P. 1980, Acta Astronomica,
             30, 1

\reference{} Agol, E., \& Krolik, J. H. 2000, ApJ, 528, 161

\reference{} Armitage, P. J., Reynolds, C. S., \& Chiang, J., 
             astro-ph/0007042

\reference{} Bardeen, J. M. 1970, Nature, 226, 64

\reference{} Bardeen, J. M., Carter, B., \& Hawking, S. W. 1973,
             Commun. Math. Phys., 31, 161

\reference{} Bardeen, J. M., Press, W. H., and Teukolsky, S. A. 1972,
             ApJ, 178, 347

\reference{} Begelman, M. C., Blandford, R. D., \& Rees, M. J. 1984,
             Rev. Mod. Phys., 56, 255.

\reference{} Blandford, R. D. 1999, in Astrophysical Disks,  eds. 
             J. A. Sellwood \& J. Goodman, ASP Conf. Series,
             160, 265

\reference{} Blandford, R. D., \& Znajek, R. L. 1977, MNRAS, 179, 433

\reference{} Christodoulou, D. 1970, Phys. Rev. Lett., 25, 1596

\reference{} Gammie, C. F. 1999, ApJ, 522, L57

\reference{} Gruzinov, A. 1999, astro-ph/9908101

\reference{} Krolik, J. H. 1999, ApJ, 515, L73

\reference{} Lee, H. K., Wijers, R. A. M. J., \& Brown, G. E. 2000,
             Phys. Rep., 325, 83

\reference{} Li, L. -X. 2000, ApJ, 533, L115

\reference{} Li, L. -X., \& Paczy\'nski, B. 2000, ApJ, 534, L197

\reference{} Macdonald, D., \& Thorne, K. S. 1982, MNRAS, 198, 345

\reference{} Meier, D. L. 1999, ApJ, 522, 753

\reference{} M\'esz\'aros, P., \& Rees, M. J. 1997, ApJ, 482, L29

\reference{} Misner, C. W., Thorne, K. S., \& Wheeler, J. A. 1973,
             Gravitation (San Francisco: W. H. Freeman and Company)

\reference{} Paczy\'nski, B. 1993,  in Relativistic Astrophysics and Particle
             Cosmology, ed. C. W. Akerlof \& M. A. Srednicki, Ann. NY
             Acad.  Sci., Vol. 688, 321

\reference{} Paczy\'nski, B., astro-ph/0004129

\reference{} Penrose, R. 1969, Rev. del Nuovo Cimento, 1, 252

\reference{} Phinney, E. S. 1983, in Astrophysical Jets, ed. A. Ferrari \& A. 
             G. Pacholczyk (Dordrecht: D. Reidel Publishing Co.), 201

\reference{} Rees, M. J., Begelman, M. C., Blandford, R. D., \& Phinney, E. S.
             1982, Nature 295, 17

\reference{} Thorne, K. S. 1974, ApJ, 191, 507

\reference{} Wald, R. M. 1974, ApJ, 191, 231

\end{references}
\end{document}